\documentclass[11pt,letterpaper,fleqn]{article}
\usepackage{graphicx}
\usepackage{lineno}
\usepackage{hyperref}
\hypersetup{
  colorlinks=true,
  urlcolor=blue,
  citecolor=blue
}
\usepackage{subfigure}
\usepackage{cite}
\usepackage{xcolor,colortbl}
\usepackage{wrapfig}
\usepackage{enumitem} %%% uncommenting this line and associated
\selectcolormodel{rgb}
\definecolor{LHCb dark}{rgb}{0.0000,0.3412,0.6549}%From LHCb logo
\definecolor{UC red}{rgb}{0.8196,0.1176,0.2314} % from logo
\definecolor{brickred}{rgb}{0.8, 0.25, 0.33}
\definecolor{Gray}{gray}{0.85}

\let\oldenumerate\itemize
\renewcommand{\itemize}{
  \oldenumerate
  \setlength{\itemsep}{1pt}
  \setlength{\parskip}{1pt}
  \setlength{\parsep}{1pt}
}

\def\s2i2{$S^2 I^2$}

\begin{document}
%\linenumbers
\thispagestyle{empty}

\begin{flushright}
HSF-CWP-2017-14
\end{flushright}

\Large
\begin{center}
{\bf Software trigger and event reconstruction: \\ Executive Summary}
\end{center}
\vskip 1cm

\normalsize

\hangindent=1cm
{\bf Abstract:} Realizing the physics programs of the planned and upgraded high-energy physics (HEP) experiments over the next 10 years will require the HEP community to address a number of challenges in the area of software and computing. For this reason, the HEP software community has engaged in a planning process over the past two years, with the objective of identifying and prioritizing the research and development required to enable the next generation of HEP detectors to fulfill their full physics potential. The aim is to produce a Community White Paper (CWP)~\cite{HSF2017} which will describe the community strategy and a roadmap for software and computing research and development in HEP for the 2020s. The topics of event reconstruction and software triggers were considered by a joint working group and are summarized together in this document.

\vskip 1cm

\hangindent=1cm
{\bf Editors}: Vladimir Vava Gligorov$^{12,a}$ and David Lange$^{20}$
%, LPNHE, Universite Pierre et Marie Curie, Universite Paris Diderot, CNRS/IN2P3 (vgligoro@lpnhe.in2p3.fr) and David Lange, Princeton University (David.Lange@cern.ch) 

\vskip 0.2cm
\hangindent=1cm
{\bf Contributors}:
Albrecht, Johannes$^{1}$;
Bloom, Kenneth$^{2}$;
Boccali, Tommaso$^{3}$;
Boveia, Antonio$^{4,c}$;
De Cian, Michel$^{5}$;
Doglioni, Caterina$^{6,b}$;
Dziurda, Agnieszka$^{7}$;
Farbin, Amir$^{8}$;
Fitzpatrick, Conor$^{9}$;
Gaede, Frank$^{10}$;
George, Simon$^{11}$;
Gligorov, Vladimir$^{12,a}$;
Grasland, Hadrien$^{13}$;
Grillo, Lucia$^{14}$;
Hegner, Benedikt$^{7}$;
Kalderon, William$^{6}$;
Kama, Sami$^{15}$;
Koppenburg, Patrick$^{16}$;
Krutelyov, Slava$^{17}$;
Kutschke, Rob$^{18}$;
Lampl, Walter$^{19}$;
Lange, David$^{20}$;
Moyse, Ed$^{21}$;
Norman, Andrew$^{18}$;
Petric, Marko$^{7}$;
Polci, Francesco$^{12}$;
Potamianos, Karolos$^{10}$;
Ratnikov, Fedor$^{22}$;
Raven, Gerhard$^{16}$;
Ritter, Martin$^{23}$;
Rizzi, Andrea$^{3}$;
Rodrigues, Eduardo$^{24}$;
Rousseau, David$^{13}$;
Salzburger, Andy$^{7}$;
Sexton Kennedy, Liz$^{18}$;
Sokoloff, Michael D$^{24}$;
Stewart, Graeme$^{7}$;
Ustyuzhanin, Andrey$^{22}$;
Viren, Brett$^{25}$;
Williams, Mike$^{26}$;
Winklmeier, Frank$^{27}$;
Wuerthwein, Frank$^{17}$
\bigskip 
\par {\footnotesize $^{1}$ Technische Universit\"at Dortmund, Dortmund, Germany}
\par {\footnotesize $^{2}$ University of Nebraska-Lincoln, Lincoln, NE, USA}
\par {\footnotesize $^{3}$ INFN Sezione di Pisa, Università di Pisa, Scuola Normale Superiore di Pisa, Pisa, Italy}
\par {\footnotesize $^{4}$ The Ohio State University, Columbus, OH, USA}
\par {\footnotesize $^{5}$ Physikalisches Institut, Ruprecht-Karls-Universitat Heidelberg, Heidelberg, Germany}
\par {\footnotesize $^{6}$ Fysiska institutionen, Lunds Universitet, Lund, Sweden}
\par {\footnotesize $^{7}$ CERN, Geneva, Switzerland}
\par {\footnotesize $^{8}$ Department of Physics, The University of Texas at Arlington, Arlington, TX, USA}
\par {\footnotesize $^{9}$ Institute of Physics, École Polytechnique F\'ed\'erale de Lausanne (EPFL), Lausanne, Switzerland}
\par {\footnotesize $^{10}$ Deutsches Elektronen-Synchrotron, Hamburg, Germany}
\par {\footnotesize $^{11}$ Department of Physics, Royal Holloway University of London, Surrey, United Kingdom}
\par {\footnotesize $^{12}$ LPNHE, Université Pierre et Marie Curie, Université Paris Diderot, CNRS/IN2P3, Paris, France}
\par {\footnotesize $^{13}$ LAL, Université Paris-Sud and CNRS/IN2P3, Orsay, France}
\par {\footnotesize $^{14}$ School of Physics and Astronomy, University of Manchester, Manchester, United Kingdom}
\par {\footnotesize $^{15}$ Physics Department, Southern Methodist University, Dallas TX, United States of America}
\par {\footnotesize $^{16}$ Nikhef National Institute for Subatomic Physics and Vrije Universiteit Amsterdam, Amsterdam, The Netherlands}
\par {\footnotesize $^{17}$ University of California, San Diego, La Jolla, CA, USA}
\par {\footnotesize $^{18}$ Fermi National Accelerator Laboratory, Batavia, IL, USA}
\par {\footnotesize $^{19}$ Department of Physics, University of Arizona, Tucson, AZ, USA}
\par {\footnotesize $^{20}$ Princeton University, Princeton, NJ, USA}
\par {\footnotesize $^{21}$ Department of Physics, University of Massachusetts, Amherst, MA, USA}
\par {\footnotesize $^{22}$ National Research University Higher School of Economics, Moscow, Russia;  Yandex School of Data Analysis, Moscow, Russia}
\par {\footnotesize $^{23}$ Fakult\"at f\"ur Physik, Ludwig-Maximilians-Universit\"at M\"unchen, M\"unchen, Germany}
\par {\footnotesize $^{24}$ University of Cincinnati, Cincinnati, OH, USA}
\par {\footnotesize $^{25}$ Physics Department, Brookhaven National Laboratory, Upton, NY, USA}
\par {\footnotesize $^{26}$ Laboratory for Nuclear Science, Massachusetts Institute of Technology, Cambridge, MA, USA}
\par {\footnotesize $^{27}$ Center for High Energy Physics, University of Oregon, Eugene, OR, USA}
\bigskip
\par {\footnotesize $^{a}$ Vladimir V. Gligorov acknowledges funding from the European Research Council (ERC) under the European Union's Horizon 2020 research and innovation programme under grant agreement No 724777 “RECEPT”}
\par {\footnotesize $^{b}$ Caterina Doglioni acknowledges funding from the European Research Council (ERC) under the European Union's Horizon 2020 research and innovation programme under grant agreement No 679305 “DARKJETS”}
\par {\footnotesize $^{c}$ Antonio Bovela acknowledges funding for US Department of Energy (Grant DE-SC0011726).}

\newpage
%\tableofcontents
%\newpage
\setcounter{page}{1}

\section{Introduction}

Realizing the physics programs of the planned and/or upgraded high-energy physics (HEP) experiments over the next 10 years will require the HEP community to address a number of challenges in the area of software and computing. For this reason, the HEP software community has engaged in a planning process over the past two years, with the objective of identifying and prioritizing the research and development required to enable the next generation of HEP detectors to fulfill their full physics potential. The aim is to produce a Community White Paper (CWP)~\cite{HSF2017} which will describe the community strategy and a roadmap for software and computing research and development in HEP for the 2020s. This activity is organized under the umbrella of the HEP Software Foundation (HSF). 

The CWP process was carried out by working groups centered on specific topics. The topics of event reconstruction and software triggers are summarized together in this document and in more detail elsewhere~\cite{RECOCWP}. The reconstruction of raw detector data and simulated data and its processing in real time represent a major component of today's computing requirements in HEP. A recent projection [Campana2016] of the ATLAS 2016 computing model results in $>$85\% of the HL-LHC CPU resources being spent on the reconstruction of data or simulated events. We have evaluated the most important components of next generation algorithms, data structures, and code development and management paradigms needed to cope with highly complex environments expected in HEP detector operations in the next decade. New approaches to data processing were also considered, including the use of novel, or at least, novel to HEP, algorithms, and the movement of data analysis into real-time environments. 

This document will discuss software algorithms essential to the interpretation of raw detector data into analysis-level objects. Specifically, these algorithms can be broadly grouped: 
\begin{enumerate}[itemsep=0ex]
\item
{\bf Online}: Algorithms, or sequences of algorithms, executed on events read out from the detector in near-real-time as part of the software trigger, typically on a computing facility located close to the detector itself.
\item
{\bf Offline}: As distinguished from online, any algorithm or sequence of algorithms executed on the subset of events preselected by the trigger system, or generated by a Monte Carlo simulation application, typically in a distributed computing system.
\item
{\bf Reconstruction}: The transformation of raw detector information into higher level objects used in physics analysis. A defining characteristic of “reconstruction” which separates it from “analysis” is that the quality criteria used in the reconstruction to, for example, minimize the number of fake tracks, should be general enough to be used in the full range of physics studies required by the experimental physics program. This usually implies that reconstruction algorithms use the entirety of the detector information to attempt to create a full picture of each interaction in the detector. Reconstruction algorithms are also typically run as part of the processing carried out by centralized computing facilities.
\item
{\bf Trigger}: the part of the online system responsible for classification of events which reduces either the number of events which are kept for further “offline” analysis, the size of such events, or both. In this working group we were only concerned with software triggers, whose defining characteristic is that they process data without a fixed latency. Software triggers are part of the real-time processing path and must make decisions quickly enough to keep up with the incoming data, possibly using substantial disk buffers.
\item
{\bf Real-time analysis}: Data processing that goes beyond object reconstruction, and is performed online within the trigger system. The typical goal of real-time analysis is to combine the products of the reconstruction algorithms (tracks, clusters, jets...) into complex objects (hadrons, gauge bosons, new physics candidates...) which can then be used directly in analysis without an intermediate reconstruction step. 
\end{enumerate}

\section{Challenges}

Software trigger and event reconstruction techniques in HEP face a number of new challenges in the next decade. These are broadly categorized into 1) those from new and upgraded accelerator facilities, 2) from detector upgrades and new detector technologies, 3) increases in anticipated event rates to be processed by algorithms (both online and offline), and 4) from evolutions in software development practices.

Advancements in facilities and future experiments bring a dramatic increase in physics reach, as well as increased event complexity and rates. At the HL-LHC, the central challenge for object reconstruction is thus to maintain excellent efficiency and resolution in the face of high pileup values, especially at low transverse momenta. Detector upgrades such as increases in channel density, high precision timing and improved detector geometric layouts are essential to overcome these problems. For software, particularly for triggering and event reconstruction algorithms, there is a critical need not to dramatically increase the processing time per event.

A number of new detector concepts are proposed on the 5-10 year timescale in order to help in overcoming the challenges identified above. In many cases, these new technologies bring novel requirements to software trigger and event reconstruction algorithms or require new algorithms to be developed. Ones of particular importance at the HL-LHC include high-granularity calorimetry, precision timing detectors, and hardware triggers based on tracking information which may seed later software trigger and reconstruction algorithms. Longer term projects with sufficiently mature software infrastructure can include cost implications of the simulation and reconstruction algorithms into the detector design considerations. This is especially important when the computing cost is expected to be a substantial part of the total construction and operation cost for an experiment. 

Trigger systems for next-generation experiments are evolving to be more capable, both in their ability to select a wider range of events of interest for the physics program of their experiment, and their ability to stream a larger rate of events for further processing.  ATLAS and CMS both target systems where the output of the hardware trigger system is increased by 10x over the current capability, up to 1 MHz~\cite{ATLAS2015,CMS2015}. In other cases, such as LHCb~\cite{LHCb2014} and ALICE~\cite{ALICE2015}, the full collision rate (between 30 to 40 MHz for typical LHC operations) will be streamed to real-time or quasi-realtime software trigger systems. The increase in event complexity also brings a “problem” of overabundance of signal to the experiments, and specifically the software trigger algorithms. The evolution towards a genuine real-time analysis of data has been driven by the need to analyze more signal than can be written out for traditional processing, and technological developments which make it possible to do this without reducing the analysis sensitivity or introducing biases.

Evolutions in computing technologies are both opportunities to move beyond commodity x86 technologies, which HEP has used very effectively over the past 20 years, and significant challenges to derive sufficient event processing throughput per cost to reasonably enable our physics programs~\cite{Bird2014}.  Specific items identified included 1) the increase of SIMD capabilities (processors capable of running a single instruction set simultaneously over multiple data), 2) the evolution towards multi- or many-core architectures, 3) the slow increase in memory bandwidth relative to CPU capabilities, 4) the rise of heterogeneous hardware, and 5) the possible evolution in facilities available to HEP production systems. 

The move towards open source software development and continuous integration systems brings opportunities to assist developers of software trigger and event reconstruction algorithms. Continuous integration systems have already allowed automated code quality and performance checks, both for algorithm developers and code integration teams. Scaling these up to allow for sufficiently high statistics checks is among the still outstanding challenges. As the timescale for experimental data taking and analysis increases, the issues of legacy code support increase. Code quality demands increase as traditional offline analysis components migrate into trigger systems, or more generically into algorithms that can only be run once. 

\section{Current approaches}
  
Substantial computing facilities are in use for both online and offline event processing across all experiments surveyed. Online facilities are dedicated to the operation of the software trigger, while offline facilities are shared for operational needs including event reconstruction, simulation (often the dominant component) and analysis. CPU in use by experiments is typically at the scale of tens or hundreds of thousands of x86 processing cores. Projections to future needs, such as for the HL-LHC, show the need for a substantial increase in scale of facilities without significant changes in approach or algorithms.

The CPU needed for event reconstruction tends to be dominated by charged particle reconstruction (tracking), especially when the need for efficiently reconstructing low transverse momentum particles is considered. Calorimetric reconstruction, particle flow reconstruction, particle identification algorithms also make up significant parts of the CPU budget in some experiments. The CPU required for event reconstruction and trigger area with challenges and substantial potential risk to the computing cost of experiments. In this respect, software for future experiments will continue to evolve, to improve both the physics and technical performance characteristics of algorithms, and the uncertainty and evolution due to detector performance and operating conditions will continue throughout the experimental program.

Disk storage is typically 10s to 100s of PB per experiment. It is dominantly used to make the output of the event reconstruction, both for real data and simulation, available for analysis. Current generation experiments have moved towards smaller, but still flexible, data tiers for analysis. These tiers are typically based on the ROOT~\cite{Brun1996} file format and constructed to facilitate both skimming of interesting events and the selection of interesting pieces of events by individual analysis groups or through centralized analysis processing systems. Initial implementations of real-time analysis systems are in use within several experiments. These approaches remove the detector data that typically makes up the raw data tier kept for offline reconstruction, and keep only final analysis objects~\cite{Aaij2016,ATLAS2017,CMS2016}.

Detector calibration and alignment requirements were surveyed. Generally a high level of automation is in place across experiments, both for very frequently updated measurements and more rarely updated measurements. Often automated procedures are integrated as part of the data taking and data reconstruction processing chain. Some longer term measurements, requiring significant data samples to be analyzed together remain as critical pieces of calibration and alignment work. These techniques are often most critical for a subset of precision measurements rather than for the entire physics program of an experiment.

\section{Research and development Roadmap and Goals}  

We identify seven broad areas to be critical for software trigger and event reconstruction work over the next decade. These are:
\begin{enumerate}[itemsep=-1ex]
\item
Enhanced vectorization programming techniques
\item
Algorithms and data structures to efficiently exploit many-core architectures
\item
Algorithms and data structures for non-x86 architectures (e.g., GPUs, FGAs)
\item
Enhanced quality assurance (QA) and quality control (QC) for reconstruction techniques
\item
Real-time analysis
\item
Precision physics-object reconstruction, identification and measurement techniques
\item
Fast software trigger and reconstruction algorithms for high-density environments
\end{enumerate}
Not all roadmap areas are directly applicable to the event reconstruction and triggering approach taken by all experiments. However, we expect that each area of proposed research and development will be broadly applicable to future high-energy physics experimental programs.

\subsection*{Roadmap area 1: Enhanced vectorization programming techniques} 

HEP developed toolkits and algorithms typically make poor use of vector units on commodity computing systems. Improving this will bring speedups to applications running on both current computing systems and most future architectures. The goal for work in this area is to evolve current toolkit and algorithm implementations, and best programming techniques to better use SIMD capabilities of current and future computing architectures.

\subsection*{Roadmap area 2: Algorithms and data structures to efficiently exploit many-core architectures}
  
Computing platforms are generally evolving towards having more cores in order to increase processing capability. This evolution has resulted in multi-threaded frameworks in use, or in development, across HEP. Algorithm developers can improve throughput by being thread safe and enabling the use of fine-grained parallelism. The goal is to evolve current event models, toolkits and algorithm implementations, and best programming techniques to improve the throughput of multithreaded software trigger and event reconstruction applications.

\subsection*{Roadmap area 3: Algorithms and data structures for non-x86 computing architectures (e.g., GPUs, FPGAs)}

Computing architectures using technologies beyond CPUs offer an interesting alternative for increasing throughput of the most time consuming trigger or reconstruction algorithms. Such architectures (e.g., GPUs, FPGAs) could be easily integrated into dedicated trigger or specialized reconstruction processing facilities (e.g., online computing farms). The goal is to demonstrate how the throughput of toolkits or algorithms can be improved through the use of new computing architectures in a production environment. 

\subsection*{Roadmap area 4: Enhanced QA/QC for reconstruction techniques}

HEP experiments have extensive continuous integration systems, including varying code regression checks that have enhanced the quality assurance and quality control procedures for software development in recent years. These are typically maintained by individual experiments and have not yet reached the scale where statistical regression, technical, and physics performance checks can be performed for each proposed software change. The goal is to enable the development, automation, and deployment of extended QA and QC tools and facilities for software trigger and event reconstruction algorithms. 

\subsection*{Roadmap area 5: Real-time analysis }

Real-time analysis techniques are being adopted to enable a wider range of physics signals to be saved by the trigger for final analysis. As rates increase, these techniques can become more important and widespread by enabling only the parts of an event associated with the signal candidates to be saved, reducing the required disk space. The goal is to evaluate and demonstrate the tools needed to facilitate real-time analysis techniques. Research topics include compression and custom data formats; toolkits for real-time detector calibration and validation which will enable full offline analysis chains to be ported into real-time; and frameworks which will enable non-expert offline analysts to design and deploy real-time analyses without compromising data taking quality. 

\subsection*{Roadmap area 6: Precision physics-object reconstruction, identification and measurement techniques}

The central challenge for object reconstruction at HL-LHC is to maintain excellent efficiency and resolution in the face of high pileup values, especially at low transverse momenta. Both trigger and reconstruction approaches need to exploit new techniques and higher granularity detectors to maintain or even improve physics measurements in the future. Reconstruction in very high pileup environments, such as the HL-LHC or FCC-hh, may also greatly benefit from adding timing information to our detectors, in order to exploit the finite beam crossing time during which interactions are produced. The goal is to develop and demonstrate efficient techniques for physics object reconstruction and identification in complex environments.

\subsection*{Roadmap area 7: Fast software trigger and reconstruction algorithms for high-density environments }

Future experimental facilities will bring a large increase in event complexity. The scaling of current-generation algorithms with this complexity must be improved to avoid a large increase in resource needs. In addition, it may be desirable or indeed necessary to deploy new algorithms, including advanced machine learning techniques developed in other fields, in order to solve these problems. The goal is to evolve or rewrite existing toolkits and algorithms focused on their physics and technical performance at high event complexity (e.g. high pileup at HL-LHC). Most important targets are those which limit expected throughput performance at future facilities (e.g., charged-particle tracking). A number of such efforts are already in progress across the community.

\section{Conclusions}

The next decade will see the volume and complexity of data being processed by HEP experiments increase by at least one order of magnitude. While much of this increase is driven by planned upgrades to the four major LHC detectors, new experiments such as DUNE will also make significant demands on the HEP data processing infrastructure. It is essential that software triggers and event reconstruction algorithms continue to evolve so that they are able to efficiently exploit future computing architectures and deal with this increase in data rates without loss of physics capability. 

We have identified seven key areas where R\&D is necessary to enable the community to exploit the full power of the enormous datasets which we will be collecting. Three of these areas concern the increasingly parallel and heterogeneous computing architectures which we will have to write our code for. In addition to a general effort to vectorize our codebases, we must understand what kinds of algorithms are best suited to what kinds of hardware architectures, develop benchmarks that allow us to compare the physics-per-dollar-per-watt performance of different algorithms across a range of potential architectures, and find ways to optimally utilise heterogeneous processing centres. The consequent increase in the complexity and diversity of our codebase will necessitate both a determined push to educate tomorrow’s physicists in modern coding practices, and a development of more sophisticated and automated quality assurance and control for our codebases. The increasing granularity of our detectors, and the addition of timing information to help cope with the extreme pileup conditions at the HL-LHC, will require us to both develop new kinds of reconstruction algorithms and to make them fast enough for use in real-time. Finally, the increased signal rates will mandate a push towards real-time analysis in many areas of HEP, in particular those with low transverse momentum signatures.

The success of this R\&D program will be intimately linked to challenges confronted in other areas of HEP computing, most notably the development of software frameworks which are able to support heterogeneous parallel architectures, including the associated data structures and I/O, the development of lightweight detector models which maintain physics precision with minimal timing and memory consequences for the reconstruction, enabling the use of offline analysis toolkits and methods within real-time analysis, and the ability to integrarte machine learning reconstruction algorithms being developed outside HEP into our workflows and apply them to our problems. For this reason perhaps the most important task ahead of us is to maintain the community which has coalesced together in this CWP process, so that the work done in these sometimes disparate areas of HEP fuses coherently together into a solution to the problems facing us over the next decade.

%\section*{Acknowledgements}
%Vladimir V. Gligorov acknowledges funding from the European Research Council (ERC) under the European Union's Horizon 2020 research and innovation programme under grant agreement No 724777 ``RECEPT''. Caterina Doglioni acknowledges funding from the European Research Council (ERC) under the European Union's Horizon 2020 research and innovation programme under grant agreement No 679305  ``DARKJETS''.  Eduardo Rodrigues acknowledges support by the United States National Science Foundation under award ACI-1450319. 

%\newpage 
%\bibliographystyle{unsrt}
%\bibliography{recoCWP}

\end{document}